\documentclass[fleqn,usenatbib]{mnras}

\usepackage[T1]{fontenc}
\usepackage{ae,aecompl}
\usepackage[normalem]{ulem}

\usepackage{graphicx}	
\usepackage{amsmath}	
\usepackage{amssymb}	
\usepackage{etoolbox}
\newcommand{\code}[1]{\texttt{#1}}

\newcommand{\mesa}{\code{MESA}}

\newcommand{\mppnp}{\code{mppnp}} 

\input{vectors}
\usepackage[usenames,dvipsnames]{color}
\usepackage[]{color}

\newcommand{\arx}[1]{arXiv:1809.03666v1 [astro-ph.SR]}
\newcommand{\jphg}[1]{Journal of Physics G}

\title[An evidence-based assumption and $^7$Be in novae]
{An evidence-based assumption that helps to reduce the discrepancy between the observed and predicted $^7$Be abundances in novae}

\author[P. A. Denissenkov et al.]{Pavel A. Denissenkov,$^{1,2,3,\dagger}$\thanks{E-mail: pavelden@uvic.ca}
Chris Ruiz$^{2,3}$, Sriteja Upadhyayula$^{3}$ and Falk Herwig$^{1,2,\dagger}$
\\
$^{1}$Department of Physics \& Astronomy, University of Victoria, Victoria, B.C., V8W~2Y2, Canada\\
$^{2}$Joint Institute for Nuclear Astrophysics, Center for the Evolution of the Elements, Michigan State University, 640 South Shaw Lane,\\
   East Lansing, MI 48824, USA\\
$^{3}$TRIUMF, 4004 Wesbrook Mall,  Vancouver, BC, V6T~2A3, Canada\\
$^\dagger$NuGrid Collaboration, \href{http://nugridstars.org}{http://nugridstars.org}\\}

\date{Accepted XXX. Received YYY; in original form ZZZ}

\pubyear{2020}

\begin{document}
\label{firstpage}
\pagerange{\pageref{firstpage}--\pageref{lastpage}}
\maketitle

\begin{abstract}
Recent spectroscopic measurements of the equivalent widths of the resonant Be\,II doublet and Ca\,II K lines and their ratios in
expanding nova ejecta indicate surprisingly high abundances of $^7$Be with a typical
mass fraction $X_\mathrm{obs}(^7\mathrm{Be}) = 10^{-4}$. This is an order of magnitude larger than
theoretically predicted values of $X_\mathrm{theor}(^7\mathrm{Be})\sim 10^{-5}$ for novae. 
We use an analytical solution of the $^7$Be production equations to demonstrate that $X_\mathrm{theor}(^7\mathrm{Be})$ 
is proportional to the $^4$He mass fraction $Y$
in the nova accreted envelope and then we perform computations of 1D hydrostatic evolution of the $1.15\,M_\odot$ CO nova model that
confirm our conclusion based on the analytical solution. Our assumption of enhanced $^4$He abundances 
helps to reduce, although not completely eliminate, the discrepancy between $X_\mathrm{obs}(^7\mathrm{Be})$ and $X_\mathrm{theor}(^7\mathrm{Be})$.
It is supported by UV, optical and IR spectroscopy data that reveal unusually high values of $Y$ in nova ejecta. We also show that
a significantly increased abundance of $^3$He in nova accreted envelopes does not lead to higher values of
$X_\mathrm{theor}(^7\mathrm{Be})$ because this assumption affects the evolution of nova models
resulting in a decrease of both their peak temperatures and accreted masses and, as a consequence, in a reduced production of $^7$Be.
\end{abstract}




\section{Introduction}
\label{sec:intro}
It had been suggested a long time ago that novae could produce significant amounts of $^7$Li \citep{starrfield:78}.
This hypothesis has recently been supported by direct spectroscopic detection and measurement of high abundances of $^7$Be
in expanding ejecta of seven novae, taking into account that $^7$Li is a decay product (100\% electron capture) of
$^7$Be whose terrestrial half-life is 53.2 days \citep{tajitsu:15,tajitsu:16,molaro:16,selvelli:18,izzo:18,molaro:20}. 
However, a problem arises when we compare the observed $^7$Be mass fractions
in the novae with their theoretically predicted values and find that, on average, the former, $X_\mathrm{obs}(^7\mathrm{Be})\approx 10^{-4}$ \citep{molaro:20},
exceed the latter, $X_\mathrm{theor}(^7\mathrm{Be})\sim 10^{-5}$ \citep[e.g.,][]{hernanz:96,jose:98,starrfield:20}, by an order of magnitude.

The discrepancy between the observed and predicted $^7$Be abundances in novae can obviously be reduced either by
casting doubt on the spectroscopically derived $^7$Be abundances, showing that they were overestimated, or
by tuning up nova models under reasonable assumptions, so that they can predict higher $^7$Be abundances.
\cite{chugai:20} have explored the first possibility criticizing the assumption made in the cited
observational works that the ionic number density ratio $N(\mathrm{Be\,II})/N(\mathrm{Ca\,II})$, obtained
by comparing equivalent widths of the resonant Be\,II doublet and Ca\,II K lines
to derive $X_\mathrm{obs}(^7\mathrm{Be})$, is equal to the total number ratio $N(\mathrm{Be})/N(\mathrm{Ca})$.
\cite{chugai:20} have used observational data of \cite{molaro:16} for the nova V5668 Sgr to construct a simple model of 
its expanding photosphere and atmosphere with which they have demonstrated that it was a wrong assumption.
According to their model, the ionization fraction $N(\mathrm{Be\,II})/N(\mathrm{Be})$ is $\sim$\,10 to $\sim$\,100 times
larger than $N(\mathrm{Ca\,II})/N(\mathrm{Ca})$, which should decrease the abundance of $^7$Be in
this nova, $X_\mathrm{obs}(^7\mathrm{Be})\approx 7.3\times 10^{-4}$, by the corresponding factors,
bringing it into an agreement with theoretical predictions.

It is beyond the scope of our paper to discuss if the conclusion about a significant overestimate of the $^7$Be abundance
in V5668 Sgr and possibly in other novae made by \cite{chugai:20} is true or not. We can only add that the assumption
that most of the $^7$Be and Ca atoms are in the singly ionized state is supported, according to \cite{tajitsu:16}, by the fact that
they failed to detect doubly ionized states of Fe-peak elements with second ionization potentials intermediate to those of
Be\,II (18.21 eV) and Ca\,II (11.87 eV) in the spectra of novae V5668 Sgr and V2944 Oph. Also, \cite{selvelli:18}
obtained very close estimates of the $^7$Be abundance in the nova V838 Her using four different methods, one of which
was based on the assumption that $N(^7\mathrm{Be\,II})/N(\mathrm{Mg\,II}) = N(^7\mathrm{Be})/N(\mathrm{Mg})$ and
used an equivalent width of Mg\,II whose ionization potential 15.03 eV is closer to that of Be\,II.
We leave a further discussion of the veracity of the reported anomalously high abundances of $^7$Be in novae to
stellar spectroscopists and proceed to a brief summary of previously proposed tuning-ups of nova models that result in an increase
of their predicted $^7$Be abundances.

There are three types of models that follow in detail thermonuclear runaway (TNR) and convective nucleosynthesis in novae:
1D hydrodynamic models, 1D hydrostatic stellar evolution models, and one- or two-zone parametric models,
the last ones being only used for simple estimates. For the frequently modelled CO nova with
the white-dwarf (WD) mass $1.15\,M_\odot$, the largest value of $X_\mathrm{theor}(^7\mathrm{Be}) = 1.9\times 10^{-5}$
has been obtained by \cite{starrfield:20} in their 1D hydrodynamic simulation in which the solar composition of accreted envelope
was switched to a mixture of 25\% of the WD and 75\% of the solar compositions immediately after a start of the TNR, when convective mixing
had begun, to mimic the dredge-up of the WD material by convection, as revealed by 3D hydro simulations of \cite{casanova:16}.
Our nova evolutionary model with 50\% pre-mixed WD composition in the accreted envelope computed using the MESA code implemented in the 
Nova Framework \citep{denissenkov:14} predicts the same value of $X_\mathrm{theor}(^7\mathrm{Be}) = 1.2\times 10^{-5}$ as
the one obtained by \cite{starrfield:20} for their second considered case of a mixture of equal amounts of the WD and solar compositions.
This is larger than the value of $X_\mathrm{theor}(^7\mathrm{Be}) = 8.1\times 10^{-6}$ reported by \cite{jose:98} for their
1D hydrodynamic nova model with the similar parameters.
The two-zone parametric model of \cite{chugai:20} has an upper limit of $X_\mathrm{theor}(^7\mathrm{Be}) = 3\times 10^{-5}$.

Because $^7$Be is produced in the reaction $^3$He($\alpha,\gamma)^7$Be, it has been suggested many times, but, as far as we know,
has not been verified in nova simulations yet, that a higher $^3$He mass fraction in accreted matter could help to reduce
the discrepancy between the observed and predicted $^7$Be abundances in novae.
In this Letter, we will show that, instead of $^3$He, it is rather an assumed enhanced abundance of $^4$He, which is
indeed observed in nova envelopes \citep{gehrz:98,downen:13}, that can help to push the predicted abundance of $^7$Be in novae
closer to its observed values, that will be represented here by the value of $X_\mathrm{obs}(^7\mathrm{Be}) = 10^{-4}$
considered as a typical $^7$Be yield for novae by \cite{molaro:20}, thus reducing the discrepancy between observations and theory.

\section{Equations of the $^7$Be production and their analytical solution}

The rates of the reactions that affect the production of $^7$Be in novae have negligible uncertainties, which
leaves us with a few parameters whose variations within their reasonable limits may lead to a significant increase of this production.
The surprisingly good agreement between the $^7$Be yields provided for the similar nova models by different simulations means that it is probably not
details of the physics of nova explosion that mainly determine the predicted $^7$Be abundance in its ejecta. Therefore, we have decided to
vary the initial abundance of $^3$He, as was previously proposed, and also that of $^4$He, because these parameters can still
be considered as relatively free (their possible variation ranges will be discussed later) 
and because their values are expected to directly affect the $^7$Be production in the reaction $^3$He($\alpha,\gamma)^7$Be.
The fruitful consumption of $^3$He by this reaction is accompanied by its waste in the competing reaction
$^3$He($^3$He,2{\it p})$^4$He whose cross section is five orders of magnitude larger. At given temperature $T$, density $\rho$,
and constant $^4$He mass fraction $Y$,
this competition is described by the following $^7$Be production equations:
\begin{eqnarray}
\frac{dX(^3\mathrm{He})}{dt} & = &  -\frac{1}{3}\lambda_1\rho\left[X(^3\mathrm{He})\right]^2 - \frac{1}{4}\lambda_2\rho X(^3\mathrm{He}) Y, \nonumber \\
\frac{dX(^7\mathrm{Be})}{dt} & = & \frac{7}{12}\lambda_2\rho X(^3\mathrm{He}) Y, \nonumber
\end{eqnarray}
where $\lambda_{1}$ and $\lambda_2$ are $T$-dependent rates ($\lambda_i\equiv \langle\sigma v \rangle_iN_\mathrm{A}$ is the product of
the Maxwellian-averaged cross section and Avogadro number)
of the reactions $^3$He($^3$He,2{\it p})$^4$He and $^3$He($\alpha,\gamma)^7$Be, respectively.
The analytical solution of these equations is
\begin{equation}
X(^7\mathrm{Be}) = \frac{7}{4} Y\frac{\lambda_2}{\lambda_1}\ln \left[ 1 + \frac{4}{3}\frac{X_0(^3\mathrm{He})}{Y}\frac{\lambda_1}{\lambda_2}
\left( 1 - e^{-t/\tau}\right)\right],
\label{eq:sol}
\end{equation}
where $\tau = 4/(\lambda_2\rho Y)$, and $X_0(^3\mathrm{He})$ is the initial mass fraction of $^3$He. 

\begin{figure}
  \centering
  \includegraphics[width=\columnwidth]{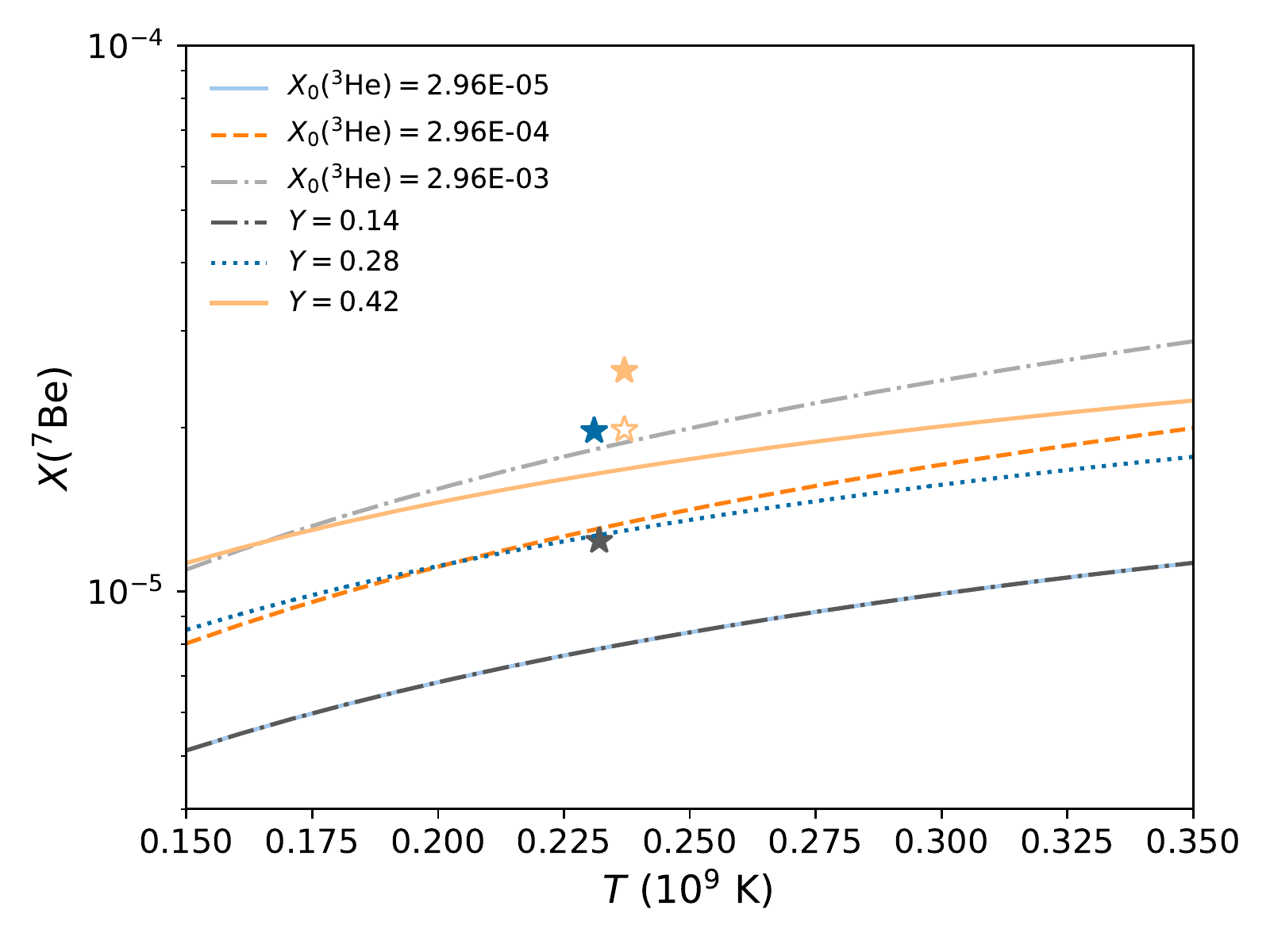}
  \caption{The mass fractions of $^7$Be as functions of temperature
       predicted by the analytical solution (\ref{eq:sol}) for different initial abundances of $^3$He and $^4$He
       (curves, when one of the abundances is varied the other is kept constant with its value taken from the mixture of the equal amounts of the WD
       and solar compositions, i.e. $X_0(^3\mathrm{He}) = 2.96\times 10^{-5}$ and $Y=0.14$). The star symbols are the $^7$Be abundances
       obtained with our $1.15\,M_\odot$ CO nova evolutionary model for the same values of $Y$ as the ones used in the analytical solutions
       represented by the curves of the same colours. The open star symbol shows the reduced $^7$Be yield in the nova model with $Y=0.42$ and 
       enhanced $X_0(^3\mathrm{He}) = 2.96\times 10^{-3}$.
  }
  \label{fig:analytical}
\end{figure}

Taking a half-solar value of $Y=0.14$, for
the density $\rho = 261\,\mathrm{g\,cm}^{-3}$ and temperature $T_9\equiv T/10^9\,\mathrm{K} = 0.232$ that are achieved
near the maxima of energy generation rates in both the CNO cycle and the pp chains at the bottom of the accreted envelope 
in our $1.15\,M_\odot$ CO nova model (Section \ref{sec:results}), we find the timescale $\tau \approx 40\,\mathrm{s}$ that is comparable
to the timescale of very fast changes of the nova $\rho$ and $T$ profiles during a time interval of $\sim$\,$100\,\mathrm{s}$ around its peak temperature.
By the end of this period, $X(^7\mathrm{Be})$ has already attained its maximum value. Therefore,
when estimating the final values of $X(^7\mathrm{Be})$ predicted by our analytical solution,
we ignore the exponent in Equation~(\ref{eq:sol}), assuming that $t/\tau\gg 1$.
These values are plotted in Figure~\ref{fig:analytical} for a range of $T$ and for different values of the parameters $X_0(^3\mathrm{He})$ and $Y$, keeping
one of them fixed at its half-solar value, i.e. $X_0(^3\mathrm{He})=2.96\times 10^{-5}$ and $Y=0.14$ 
for the 50\% WD pre-mixed accreted matter, and assuming that all $^2$H was transformed into $^3$He in the donor star, 
when changing the other. 

From Figure~\ref{fig:analytical}
we see that $X(^7\mathrm{Be})$ increases with $T$, which agrees with its increase with the WD mass found by \cite{starrfield:20}.
The range of the $^7$Be yield predicted for novae by our analytical solution for the half-solar values of both
$X_0(^3\mathrm{He})$ and $Y$ encompasses almost all the values of $X(^7\mathrm{Be})$ predicted by the corresponding 1D nova models.
It is clear from Equation~(\ref{eq:sol}) that the persistent prediction of $X(^7\mathrm{Be})\sim 10^{-5}$ by nova models is simply caused by the fact that
$\lambda_2/\lambda_1\sim 10^{-5}$ and this ratio weakly depends on $T$. 

Given that the predicted value of $X(^7\mathrm{Be})\gg X_\odot(^7\mathrm{Li})$, Equation~(\ref{eq:sol})
can also be written as        
\begin{equation}
\frac{X(^7\mathrm{Li})}{X_0(^7\mathrm{Li})} = 1 + C\log_{10}\frac{X_0(^3\mathrm{He})}{X_\odot(^3\mathrm{He})},
\label{eq:boffin}
\end{equation}
where $C = \left\{\log_{10}\left[\frac{4}{3}\frac{X_\odot(^3\mathrm{He})}{Y}\frac{\lambda_1}{\lambda_2}\right]\right\}^{-1}$,
and $X_0(^7\mathrm{Li})$ is the abundance of $^7$Li for the case of $X_0(^3\mathrm{He}) = X_\odot(^3\mathrm{He})$.
This equation assumes that $X(^7\mathrm{Li}) = X(^7\mathrm{Be})$ and again that $t/\tau\gg 1$.
It has been derived by \cite{boffin:93} with the value of $C = 1.5$ constrained by results of their parametric
one-zone nova model calculations. For their preferred value of $T_9 = 0.3$,
our analytical solution provides $C = 0.75$. The logarithmic dependence of $X(^7\mathrm{Be})$ on $X_0(^3\mathrm{He})$ is
clearly seen in Figure~\ref{fig:analytical}. 

The maximum $2\,\mathrm{dex}$ enhancement of the initial $^3$He abundance relative
to its solar value, assumed in Figure~\ref{fig:analytical} for the accreted matter, could potentially come from
the so-called $^3$He bump inside the donor star, formed as a result of incomplete H burning in the pp I branch, like in all solar-type stars.
Then, given that the WD companion is most likely to be tidally locked, its rapid rotation and tidal deformation
should drive meridional circulation that may reach the bump and bring the abundant $^3$He to the surface from where
it will be donated to the WD. It is also possible that the companion has lost enough mass to expose the $^3$He bump
at its surface \citep{shen:09}.

Equation~(\ref{eq:sol}) and Figure~\ref{fig:analytical} also show that the amount of $^7$Be produced in novae is proportional to $Y$,
hence its increase by the factors of 2 and 3 should result in the enhancements of $X(^7\mathrm{Be})$ comparable to those
obtained for $X_0(^3\mathrm{He})/X_\odot(^3\mathrm{He}) = 10\ \mathrm{and}\ 100$. Table~2 of \cite{gehrz:98} and Table~1 of \cite{downen:13} summarize
UV, optical and IR spectroscopy data on mass fractions of H, He and heavy elements in nova ejecta. Most of the $Y$ values
presented in this table exceed the half-solar value of $Y\approx 0.14$ used in our and other similar nova models with
the 50\% pre-mixed WD composition, nearly 30\% of them being larger than $0.3$ with six stars having $Y > 0.4$. Therefore, the assumption of
$Y > 0.14$ in the nova envelope that we make in this work is supported by observations.

Because $^3$He burning starts at the bottom of the accreted envelope before the TNR is triggered
by the reaction $^{12}$C({\it p},$\gamma)^{13}$N \citep[e.g.,][and references therein]{shen:09}, the impact of a significant
increase of its initial abundance on the $^7$Be production in novae can only be studied using full nova models,
since neither simple one- or two-zone parametric models nor even multi-zone post-processing nucleosynthesis
models, like the one of the Nova Framework that uses the NuGrid \mppnp\ code \citep{denissenkov:14},
take into account a feedback of this assumption on nova properties. Therefore, we have employed
the MESA code setup of the Nova Framework to perform 1D hydrostatic evolutionary
computations of our $1.15\,M_\odot$ CO nova model with increased abundances of $^4$He and $^3$He in its
accreted envelope, results of which are presented in the next section.

\section{Results of $1.15\,M_\odot$ CO nova evolutionary computations with increased abundances of $^4$He and $^3$He
in the accreted envelope} 
\label{sec:results}

Our Nova Framework has been using the revision 5329 of the \mesa\ stellar evolution code \citep{paxton:11,paxton:13} with a content of its inlist file
set up to model CO and ONe nova evolution \citep{denissenkov:14}. In this work, we have chosen the largest
of the nuclear reaction networks available in the Nova Framework \code{nova.net} with 77 species from $^1$H to $^{40}$Ca coupled by 442 reactions for which
we have used the rates from the JINA Reaclib v1.1 database \citep{cyburt:10}. This network includes
most of the reactions affecting the $^7$Be production in novae discussed by \cite{boffin:93} and \cite{hernanz:96}, except
$^7$Be($\alpha,\gamma)^{11}$C and the chain $^8$B({\it p},$\gamma)^9$C({\it e}$^+,\nu)2 ^4$He because these are much slower than
their competing reactions $^7$Be({\it p},$\gamma)^8$B and $^8$B({\it e}$^+\bar{\nu})2 ^4$He.

\begin{figure*}
  \centering
  \includegraphics[width=144mm]{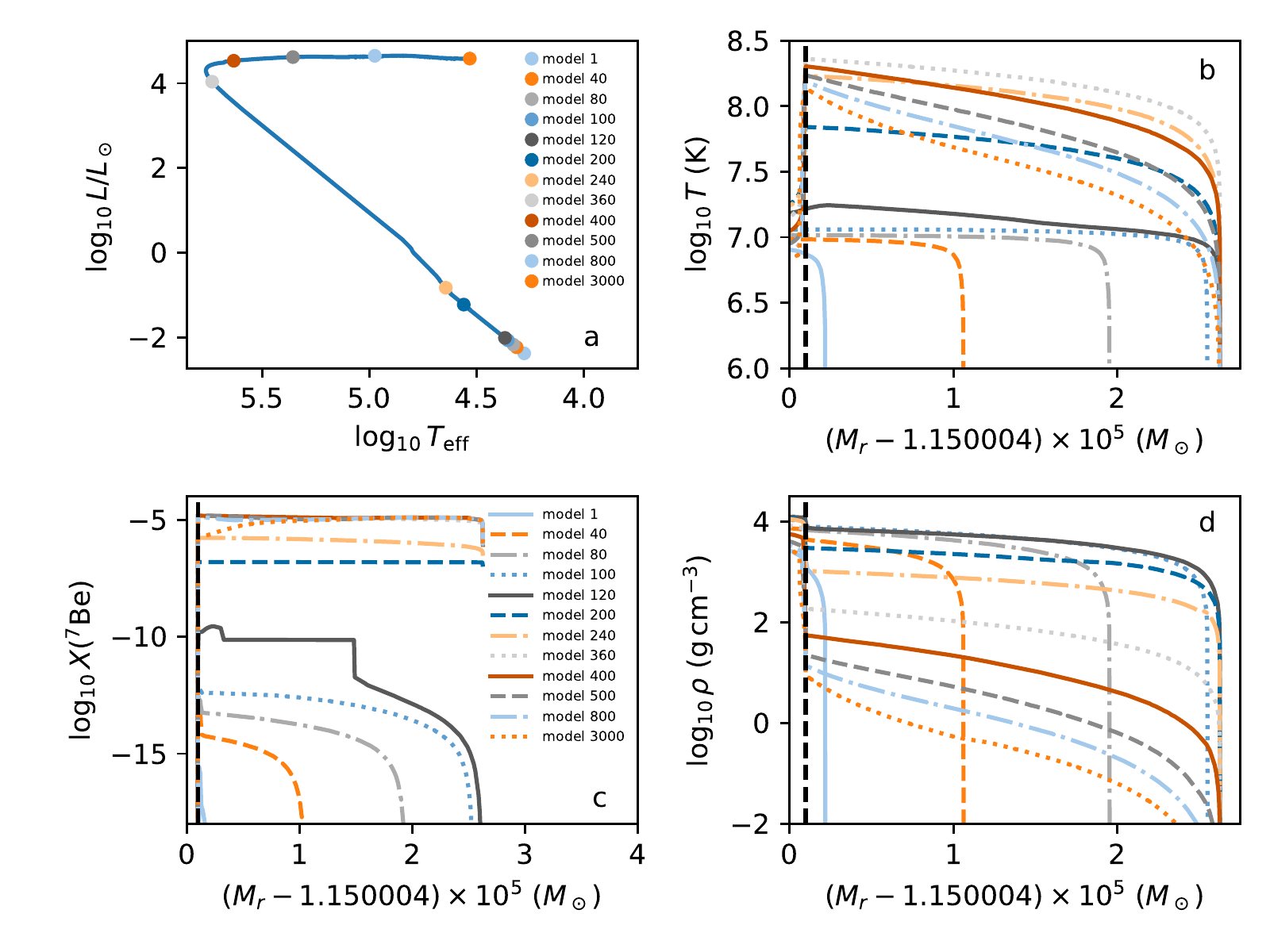}
  \caption{The track on the Hertzsprung-Russell diagram (panel a), temperature (b), $^7$Be mass fraction (c) and density (d) mass-coordinate profiles inside
  the envelope of our $1.15\,M_\odot$ CO nova model shown for different evolutionary phases indicated by model numbers.
  The $^7$Be abundance attains its maximum value between models 360 and 400, at the knee of the evolutionary track (between the light-gray
  and brown circles in panel a), when $T$ reaches its peak value at the bottom of the accreted envelope (panel b). This computation used
  the 50\% WD pre-mixed chemical composition for the envelope.
  }
  \label{fig:modelA}
\end{figure*}

Results of our computations of the evolution of the $1.15\,M_\odot$ CO nova model with the WD central temperature $12\times 10^6$ K
and accretion rate $2\times 10^{-10} M_\odot\mathrm{\,yr}^{-1}$ are presented in Figure~\ref{fig:modelA} for
the 50\% WD pre-mixed composition of the accreted envelope, including the $^4$He and $^3$He mass fractions, i.e. for
$Y=0.14$ and $X_0(^3\mathrm{He})=2.96\times 10^{-5}$. As already mentioned, we see that the $^7$Be abundance attains its maximum value of
$X_\mathrm{max}(^7\mathrm{Be}) = 1.24\times 10^{-5}$
(panel c) around the peak temperature $T_9 = 0.232$ (panel b), when the star approaches the knee of its evolutionary track 
(between the models 360 and 400 in panel a). By this moment, the density at the bottom of the convective envelope has already dropped
to $\rho = 261\ \mathrm{g\,cm}^{-3}$ from its maximum value of $8\times 10^3\ \mathrm{g\,cm}^{-3}$. 
The model has accreted $M_\mathrm{acc} = 2.53\times 10^{-5} M_\odot$
by the beginning of its TNR, and its evolution was followed for 71 minutes after the peak temperature.

After the $^4$He mass fraction in the accreted envelope was increased to the values of $Y=0.28$ and $Y=0.42$, that are
within the limits of $Y$ measured in nova ejecta \citep{gehrz:98,downen:13}, at the expense of
the $^{16}$O abundance to keep the sum of all mass fractions equal to one, the maximum $^7$Be abundances in our nova model
have reached the values of $X_\mathrm{max}(^7\mathrm{Be}) = 1.97\times 10^{-5}$ and $X_\mathrm{max}(^7\mathrm{Be}) = 2.54\times 10^{-5}$,
respectively. The other nova properties have essentially remained unchanged. All the three values of $X_\mathrm{max}(^7\mathrm{Be})$
predicted by the nova model are displayed in Figure~\ref{fig:analytical} as star symbols of the same colours that we used
to plot the analytical solution curves for the corresponding values of $Y$. Although resulting in slightly different values, unsurprisingly because we compare
the simple one-zone and detailed multi-zone evolutionary nova models,
the two sets of computations show the same trend of the predicted $^7$Be abundance increasing with $Y$.
It is interesting that our value of $X_\mathrm{max}(^7\mathrm{Be})$ obtained for $Y=0.28$ agrees very well with 
the value of $X(^7\mathrm{Be}) = 1.9\times 10^{-5}$ reported by \cite{starrfield:20} for their 1D hydrodynamic
$1.15\,M_\odot$ CO nova model with a mixture of 25\% WD and 75\% solar compositions in the envelope that has a close value of $Y=0.23$.

The results change drastically when we assume a significantly increased initial abundance of $^3$He in the accreted envelope.
For $X_0(^3\mathrm{He})/X_\odot(^3\mathrm{He}) = 100$ and $Y=0.42$ our nova model yields
$X_\mathrm{max}(^7\mathrm{Be}) = 1.98\times 10^{-5}$ which is 22\% lower than for the nova model with the same value of $Y$ and
the half-solar mass fraction of $^3$He (the filled and open star symbols of the same colour in Figure~\ref{fig:analytical}). 
The other nova properties also undergo considerable changes, e.g. the peak temperature
and the accreted mass are reduced to $T_9 = 0.168$ and $M_\mathrm{acc} = 6.72\times 10^{-6} M_\odot$.
These changes are caused by the substantially increased heating of the bottom of the accreted envelope by the energy
released in the reaction $^3$He($^3$He,2{\it p})$^4$He whose $Q$ value $12.86$ MeV is much larger than $Q=1.95$ MeV for the TNR
triggering reaction $^{12}$C({\it p},$\gamma)^{13}$N. As a result, the TNR starts earlier, with the lower peak $T$ and $M_\mathrm{acc}$. 
Thus, taking into account the feedback
of the increased value of $X_0(^3\mathrm{He})$ on the nova properties leads to a decrease, rather than to the previously expected increase, 
of the maximum $^7$Be abundance
predicted with the evolutionary nova model. Therefore, Equations (\ref{eq:sol}) and (\ref{eq:boffin}) derived for the one-zone parametric
model do not provide correct dependencies of the $^7$Be and $^7$Li nova yields on the initial abundance of $^3$He.

\section{Conclusion}

We have done 1D stellar evolution computations of the popular $1.15\,M_\odot$ CO nova model to demonstrate that the $^7$Be production in novae
increases proportionally to the $^4$He mass fraction $Y$ in their accreted envelopes, as predicted by the analytical solution (\ref{eq:sol}).
Therefore, the assumption of a 2 to 3 times enhanced value of $Y$, as compared to the half-solar value of $Y=0.14$ used in nova models accreting a mixture of
equal amounts of the WD and solar compositions, helps to reduce, although not completely eliminate, the discrepancy between the observed
and theoretically predicted $^7$Be mass fractions in novae,
$X_\mathrm{obs}(^7\mathrm{Be})\approx 10^{-4}$ and $X_\mathrm{theor}(^7\mathrm{Be})\sim 10^{-5}$.
This assumption is supported by the surprisingly high values of $Y$, with the maximum value of $Y=0.6$, reported in nova ejecta by \cite{gehrz:98}
and \cite{downen:13} that have not been explained or refuted yet. The high $^4$He abundances in nova ejecta could come from a $^4$He layer
atop of the accreting white WD that
is formed as a result of H afterburning during the post-nova supersoft X-ray phase \citep{iben:92,starrfield:98,wolf:13}.

We have also shown for the first time that the previously proposed hypothesis of a significantly enhanced, up to $2\,\mathrm{dex}$, mass fraction of
$^3$He in the nova accreted envelope does not raise $X_\mathrm{theor}(^7\mathrm{Be})$ closer to $X_\mathrm{obs}(^7\mathrm{Be})$. Instead,
it results in a decrease of $X_\mathrm{theor}(^7\mathrm{Be})$ because of the feedback of this assumption on nova properties --- the increased release of heat
in the highly energetic reaction $^3$He($^3$He,2{\it p})$^4$He at the bottom of the accreted envelope leads to an earlier TNR with lower both peak temperature
and accreted mass. 

If we extrapolate our result obtained for $Y=0.42$ (the orange star symbol in Figure~\ref{fig:analytical})
to the peak temperature $T_9 = 0.350$ of the $1.3\,M_\odot$ ONe nova model \citep[e.g., see Table~1 of][]{denissenkov:14} we will get an estimate of
$X_\mathrm{theor}(^7\mathrm{Be})\approx 3\times 10^{-5}$ that is only $\sim 3$ times smaller than the average mass fraction of $^7$Be
observed in novae, although a slight decrease of the latter is still required for a better match.

Of note for future work, the higher $^7$Be mass fraction calculated here would lead to a higher yield of 
the characteristic 478 keV gamma ray via the 10\% $^7$Be($e^{-},\nu$) branch to the 1st excited state of $^7$Li, 
as calculated in e.g. \cite{gomezgomar:98}, in which the terrestrial value of the $^7$Be EC lifetime is adopted 
under the assumption that full Be ionization is not substantial. Applying the formalism of \cite{iben:67} 
to the conditions at the end of our nova trajectory (71 minutes) implies a very small free electron capture rate, 
as expected due to the low density. Furthermore, the occupation probabilities for K-shell electrons 
under these conditions are very small, implying that most of the $^7$Be is fully-ionized at this stage and 
will not convert at the terrestrial rate to $^7$Li until conditions are met such that bound state captures 
occur frequently again. Thus the effective lifetime, considering only the small K-shell capture rate, 
is on the order of $5\times 10^6$ days. Due to the timescales involved in the nova trajectory, 
it is not clear that this variation in lifetime will affect the temporal evolution of the 478 keV line, 
and requires careful consideration of the conditions in the hot expanding ejecta. 

\section*{Acknowledgements}

FH acknowledges funding from NSERC through a Discovery Grant. This
research is supported by the National Science Foundation (USA) under
Grant No. PHY-1430152 (JINA Center for the Evolution of the Elements).

\section*{Data availability}

The data underlying this article will be shared on reasonable request to the corresponding author.




\bibliographystyle{mnras}
\bibliography{paper.bib}

\begin{thebibliography}{}
\makeatletter
\relax
\def\mn@urlcharsother{\let\do\@makeother \do\$\do\&\do\#\do\^\do\_\do\%\do\~}
\def\mn@doi{\begingroup\mn@urlcharsother \@ifnextchar [ {\mn@doi@}
  {\mn@doi@[]}}
\def\mn@doi@[#1]#2{\def\@tempa{#1}\ifx\@tempa\@empty \href
  {http://dx.doi.org/#2} {doi:#2}\else \href {http://dx.doi.org/#2} {#1}\fi
  \endgroup}
\def\mn@eprint#1#2{\mn@eprint@#1:#2::\@nil}
\def\mn@eprint@arXiv#1{\href {http://arxiv.org/abs/#1} {{\tt arXiv:#1}}}
\def\mn@eprint@dblp#1{\href {http://dblp.uni-trier.de/rec/bibtex/#1.xml}
  {dblp:#1}}
\def\mn@eprint@#1:#2:#3:#4\@nil{\def\@tempa {#1}\def\@tempb {#2}\def\@tempc
  {#3}\ifx \@tempc \@empty \let \@tempc \@tempb \let \@tempb \@tempa \fi \ifx
  \@tempb \@empty \def\@tempb {arXiv}\fi \@ifundefined
  {mn@eprint@\@tempb}{\@tempb:\@tempc}{\expandafter \expandafter \csname
  mn@eprint@\@tempb\endcsname \expandafter{\@tempc}}}

\bibitem[\protect\citeauthoryear{{Boffin}, {Paulus}, {Arnould}  \&
  {Mowlavi}}{{Boffin} et~al.}{1993}]{boffin:93}
{Boffin} H.~M.~J.,  {Paulus} G.,  {Arnould} M.,   {Mowlavi} N.,  1993, \aap,
  \href {https://ui.adsabs.harvard.edu/abs/1993A&A...279..173B} {279, 173}

\bibitem[\protect\citeauthoryear{{Casanova}, {Jos{\'e}}, {Garc{\'\i}a-Berro}
  \& {Shore}}{{Casanova} et~al.}{2016}]{casanova:16}
{Casanova} J.,  {Jos{\'e}} J.,  {Garc{\'\i}a-Berro} E.,   {Shore} S.~N.,  2016,
  \mn@doi [\aap] {10.1051/0004-6361/201628707}, \href
  {https://ui.adsabs.harvard.edu/abs/2016A&A...595A..28C} {595, A28}

\bibitem[\protect\citeauthoryear{{Chugai} \& {Kudryashov}}{{Chugai} \&
  {Kudryashov}}{2020}]{chugai:20}
{Chugai} N.~N.,  {Kudryashov} A.~D.,  2020, arXiv e-prints, \href
  {https://ui.adsabs.harvard.edu/abs/2020arXiv200707044C} {p. arXiv:2007.07044}

\bibitem[\protect\citeauthoryear{{Cyburt} et~al.,}{{Cyburt}
  et~al.}{2010}]{cyburt:10}
{Cyburt} R.~H.,  et~al., 2010, \mn@doi [\apjs] {10.1088/0067-0049/189/1/240},
  \href {https://ui.adsabs.harvard.edu/abs/2010ApJS..189..240C} {189, 240}

\bibitem[\protect\citeauthoryear{{Denissenkov} et~al.,}{{Denissenkov}
  et~al.}{2014}]{denissenkov:14}
{Denissenkov} P.~A.,  et~al., 2014, \mn@doi [\mnras] {10.1093/mnras/stu1000},
  \href {https://ui.adsabs.harvard.edu/abs/2014MNRAS.442.2058D} {442, 2058}

\bibitem[\protect\citeauthoryear{{Downen}, {Iliadis}, {Jos{\'e}}  \&
  {Starrfield}}{{Downen} et~al.}{2013}]{downen:13}
{Downen} L.~N.,  {Iliadis} C.,  {Jos{\'e}} J.,   {Starrfield} S.,  2013,
  \mn@doi [\apj] {10.1088/0004-637X/762/2/105}, \href
  {https://ui.adsabs.harvard.edu/abs/2013ApJ...762..105D} {762, 105}

\bibitem[\protect\citeauthoryear{{Gehrz}, {Truran}, {Williams}  \&
  {Starrfield}}{{Gehrz} et~al.}{1998}]{gehrz:98}
{Gehrz} R.~D.,  {Truran} J.~W.,  {Williams} R.~E.,   {Starrfield} S.,  1998,
  \mn@doi [\pasp] {10.1086/316107}, \href
  {https://ui.adsabs.harvard.edu/abs/1998PASP..110....3G} {110, 3}

\bibitem[\protect\citeauthoryear{{Gomez-Gomar}, {Hernanz}, {Jose}  \&
  {Isern}}{{Gomez-Gomar} et~al.}{1998}]{gomezgomar:98}
{Gomez-Gomar} J.,  {Hernanz} M.,  {Jose} J.,   {Isern} J.,  1998, \mn@doi
  [\mnras] {10.1046/j.1365-8711.1998.01421.x}, \href
  {https://ui.adsabs.harvard.edu/abs/1998MNRAS.296..913G} {296, 913}

\bibitem[\protect\citeauthoryear{{Hernanz}, {Jose}, {Coc}  \&
  {Isern}}{{Hernanz} et~al.}{1996}]{hernanz:96}
{Hernanz} M.,  {Jose} J.,  {Coc} A.,   {Isern} J.,  1996, \mn@doi [\apjl]
  {10.1086/310122}, \href
  {https://ui.adsabs.harvard.edu/abs/1996ApJ...465L..27H} {465, L27}

\bibitem[\protect\citeauthoryear{{Iben}, {Kalata}  \& {Schwartz}}{{Iben}
  et~al.}{1967}]{iben:67}
{Iben} Icko J.,  {Kalata} K.,   {Schwartz} J.,  1967, \mn@doi [\apj]
  {10.1086/149399}, \href
  {https://ui.adsabs.harvard.edu/abs/1967ApJ...150.1001I} {150, 1001}

\bibitem[\protect\citeauthoryear{{Iben}, {Fujimoto}  \& {MacDonald}}{{Iben}
  et~al.}{1992}]{iben:92}
{Iben} Icko J.,  {Fujimoto} M.~Y.,   {MacDonald} J.,  1992, \mn@doi [\apj]
  {10.1086/171171}, \href
  {https://ui.adsabs.harvard.edu/abs/1992ApJ...388..521I} {388, 521}

\bibitem[\protect\citeauthoryear{{Izzo} et~al.,}{{Izzo} et~al.}{2018}]{izzo:18}
{Izzo} L.,  et~al., 2018, \mn@doi [\mnras] {10.1093/mnras/sty435}, \href
  {https://ui.adsabs.harvard.edu/abs/2018MNRAS.478.1601I} {478, 1601}

\bibitem[\protect\citeauthoryear{{Jos{\'e}} \& {Hernanz}}{{Jos{\'e}} \&
  {Hernanz}}{1998}]{jose:98}
{Jos{\'e}} J.,  {Hernanz} M.,  1998, \mn@doi [\apj] {10.1086/305244}, \href
  {https://ui.adsabs.harvard.edu/abs/1998ApJ...494..680J} {494, 680}

\bibitem[\protect\citeauthoryear{{Molaro}, {Izzo}, {Mason}, {Bonifacio}  \&
  {Della Valle}}{{Molaro} et~al.}{2016}]{molaro:16}
{Molaro} P.,  {Izzo} L.,  {Mason} E.,  {Bonifacio} P.,   {Della Valle} M.,
  2016, \mn@doi [\mnras] {10.1093/mnrasl/slw169}, \href
  {https://ui.adsabs.harvard.edu/abs/2016MNRAS.463L.117M} {463, L117}

\bibitem[\protect\citeauthoryear{{Molaro}, {Izzo}, {Bonifacio}, {Hernanz},
  {Selvelli}  \& {della Valle}}{{Molaro} et~al.}{2020}]{molaro:20}
{Molaro} P.,  {Izzo} L.,  {Bonifacio} P.,  {Hernanz} M.,  {Selvelli} P.,
  {della Valle} M.,  2020, \mn@doi [\mnras] {10.1093/mnras/stz3587}, \href
  {https://ui.adsabs.harvard.edu/abs/2020MNRAS.492.4975M} {492, 4975}

\bibitem[\protect\citeauthoryear{{Paxton}, {Bildsten}, {Dotter}, {Herwig},
  {Lesaffre}  \& {Timmes}}{{Paxton} et~al.}{2011}]{paxton:11}
{Paxton} B.,  {Bildsten} L.,  {Dotter} A.,  {Herwig} F.,  {Lesaffre} P.,
  {Timmes} F.,  2011, \mn@doi [\apjs] {10.1088/0067-0049/192/1/3}, \href
  {https://ui.adsabs.harvard.edu/abs/2011ApJS..192....3P} {192, 3}

\bibitem[\protect\citeauthoryear{{Paxton} et~al.,}{{Paxton}
  et~al.}{2013}]{paxton:13}
{Paxton} B.,  et~al., 2013, \mn@doi [\apjs] {10.1088/0067-0049/208/1/4}, \href
  {https://ui.adsabs.harvard.edu/abs/2013ApJS..208....4P} {208, 4}

\bibitem[\protect\citeauthoryear{{Selvelli}, {Molaro}  \& {Izzo}}{{Selvelli}
  et~al.}{2018}]{selvelli:18}
{Selvelli} P.,  {Molaro} P.,   {Izzo} L.,  2018, \mn@doi [\mnras]
  {10.1093/mnras/sty2310}, \href
  {https://ui.adsabs.harvard.edu/abs/2018MNRAS.481.2261S} {481, 2261}

\bibitem[\protect\citeauthoryear{{Shen} \& {Bildsten}}{{Shen} \&
  {Bildsten}}{2009}]{shen:09}
{Shen} K.~J.,  {Bildsten} L.,  2009, \mn@doi [\apj]
  {10.1088/0004-637X/692/1/324}, \href
  {https://ui.adsabs.harvard.edu/abs/2009ApJ...692..324S} {692, 324}

\bibitem[\protect\citeauthoryear{{Starrfield}, {Truran}, {Sparks}  \&
  {Arnould}}{{Starrfield} et~al.}{1978}]{starrfield:78}
{Starrfield} S.,  {Truran} J.~W.,  {Sparks} W.~M.,   {Arnould} M.,  1978,
  \mn@doi [\apj] {10.1086/156175}, \href
  {https://ui.adsabs.harvard.edu/abs/1978ApJ...222..600S} {222, 600}

\bibitem[\protect\citeauthoryear{{Starrfield}, {Truran}, {Wiescher}  \&
  {Sparks}}{{Starrfield} et~al.}{1998}]{starrfield:98}
{Starrfield} S.,  {Truran} J.~W.,  {Wiescher} M.~C.,   {Sparks} W.~M.,  1998,
  \mn@doi [\mnras] {10.1046/j.1365-8711.1998.01312.x}, \href
  {https://ui.adsabs.harvard.edu/abs/1998MNRAS.296..502S} {296, 502}

\bibitem[\protect\citeauthoryear{{Starrfield}, {Bose}, {Iliadis}, {Hix},
  {Woodward}  \& {Wagner}}{{Starrfield} et~al.}{2020}]{starrfield:20}
{Starrfield} S.,  {Bose} M.,  {Iliadis} C.,  {Hix} W.~R.,  {Woodward} C.~E.,
  {Wagner} R.~M.,  2020, \mn@doi [\apj] {10.3847/1538-4357/ab8d23}, \href
  {https://ui.adsabs.harvard.edu/abs/2020ApJ...895...70S} {895, 70}

\bibitem[\protect\citeauthoryear{{Tajitsu}, {Sadakane}, {Naito}, {Arai}  \&
  {Aoki}}{{Tajitsu} et~al.}{2015}]{tajitsu:15}
{Tajitsu} A.,  {Sadakane} K.,  {Naito} H.,  {Arai} A.,   {Aoki} W.,  2015,
  \mn@doi [\nat] {10.1038/nature14161}, \href
  {https://ui.adsabs.harvard.edu/abs/2015Natur.518..381T} {518, 381}

\bibitem[\protect\citeauthoryear{{Tajitsu}, {Sadakane}, {Naito}, {Arai},
  {Kawakita}  \& {Aoki}}{{Tajitsu} et~al.}{2016}]{tajitsu:16}
{Tajitsu} A.,  {Sadakane} K.,  {Naito} H.,  {Arai} A.,  {Kawakita} H.,   {Aoki}
  W.,  2016, \mn@doi [\apj] {10.3847/0004-637X/818/2/191}, \href
  {https://ui.adsabs.harvard.edu/abs/2016ApJ...818..191T} {818, 191}

\bibitem[\protect\citeauthoryear{{Wolf}, {Bildsten}, {Brooks}  \&
  {Paxton}}{{Wolf} et~al.}{2013}]{wolf:13}
{Wolf} W.~M.,  {Bildsten} L.,  {Brooks} J.,   {Paxton} B.,  2013, \mn@doi
  [\apj] {10.1088/0004-637X/777/2/136}, \href
  {https://ui.adsabs.harvard.edu/abs/2013ApJ...777..136W} {777, 136}

\makeatother
\end{thebibliography}



\bsp	
\label{lastpage}
\end{document}